\documentclass[fleqn,12pt,twoside]{article}
\usepackage[headings]{espcrc1}

\readRCS
$Id: espcrc1.tex,v 1.2 2004/02/24 11:22:11 spepping Exp $
\usepackage{graphicx}
\usepackage[figuresright]{rotating}

\def\be{\begin{equation}}
\def\ee{\end{equation}}
\def\Tr{{\rm Tr}}

\newcommand{\AmS}{{\protect\the\textfont2
  A\kern-.1667em\lower.5ex\hbox{M}\kern-.125emS}}

\title{Thermal Signatures of Pairing Correlations in Nuclei and Nanoparticles}

\author{Y. Alhassid\address[MCSD]{Center for Theoretical Physics, Sloane Physics Laboratory,
Yale University, New Haven, Connecticut 06520, USA}%
        \thanks{This work was supported in part by the U.S. DOE grant No. DE-FG-0291-ER-40608.}}

\begin{document}

\maketitle

\begin{abstract}
The Bardeen-Cooper-Schrieffer (BCS) mean-field theory of the pairing interaction breaks down for nuclei and ultra-small metallic grains (nanoparticles). Finite-temperature pairing correlations in such finite-size systems can be calculated beyond the BCS theory in an auxiliary-field Monte Carlo approach. We identify thermal signatures of pairing correlations in both nuclei and nanoparticles that depend on the particle-number parity.
\end{abstract}

\section{INTRODUCTION}

The pairing interaction in bulk metals leads to a phase transition from a normal to a superconducting metal below a certain critical temperature. This phase transition was explained by the Bardeen-Cooper-Schrieffer (BCS) mean-field theory~\cite{BCS}. BCS is valid when the pairing gap $\Delta$ is much larger than the single-particle mean-level spacing $\delta$, a condition that is satisfied in bulk metals.

The pairing interaction also plays an important role in nuclei. Pairing effects in nuclei at zero temperature are well documented but much less is known about thermal signatures of pairing correlations. In nuclei $\Delta/\delta$  is of order $1$, and a typical nucleus is in the crossover regime between the bulk BCS limit and the fluctuation-dominated limit. In this crossover regime, fluctuations tend to wash out signatures of the pairing transition and an interesting question is whether pairing correlations survive despite the large fluctuations.

Experimental and theoretical studies of ultra-small metallic particles (nanoparticles) have shed light on pairing correlations in finite-size systems~\cite{vondelft01}. These ultra-small metallic grains can be connected to leads and their transport properties measured. The number of electrons on the grain is controlled by changing a gate voltage.
Nanoparticles whose linear size is below $\sim 5$ nm are also close to the fluctuations-dominated regime.

We have studied pairing correlations in nuclei beyond the BCS theory using auxiliary-field Monte Carlo (AFMC) methods in the framework of the nuclear shell model. These techniques are also known as the shell model Monte Carlo (SMMC) methods~\cite{lang93,alhassid94}. With AFMC we can carry out fully correlated and realistic calculations in much larger configuration spaces than those that can be treated by conventional methods. Recently, we have extended the AFMC methods to study pairing correlations in nanoparticles beyond the BCS approximation~\cite{afmc-nano}.

The AFMC methods are briefly discussed in Sec.~\ref{AFMC}.  Signatures of pairing correlations in the heat capacity~\cite{la01,alhassid03} and moment of inertia of nuclei~\cite{alhassid05}, calculated in the AFMC approach, are presented in Sec.~\ref{nuclei}. Analogous signatures of pairing correlations in the heat capacity and spin susceptibility of nanoparticles~\cite{afmc-nano} are presented in Sec.~\ref{nanoparticles}.

\section{AUXILIARY-FIELD MONTE CARLO (AFMC) METHODS}\label{AFMC}

Correlations beyond the mean-field approximation can be calculated by taking into account fluctuations of the mean field. This can be formally expressed by the Hubbard-Stratonovich (HS) transformation~\cite{HS}. The Gibbs ensemble $e^{-\beta H}$ at inverse temperature $\beta=1/T$ (which is also the many-body propagator in imaginary time $\beta$) can be written as a functional integral
\begin{equation}\label{eq:HS}
e^{-\beta H} = \int {\cal D}[\sigma]
G_\sigma U_\sigma \;
\end{equation}
over propagators $U_\sigma$ describing non-interacting particles in time-dependent fields $\sigma(\tau)$.  The quantity $G_\sigma$ in (\ref{eq:HS}) is a Gaussian weight.

Finite-size effects are important in both nuclei and nanoparticles. For such small systems, it is necessary to calculate thermal expectation values in the {\em canonical} ensemble in which the number of particles $A$ is fixed.  For an observable $O$
\begin{eqnarray}
\label{eq:obs} \langle O\rangle\equiv {\Tr_A(O e^{-\beta H})\over \Tr_A e^{-\beta H}}={\int
{\cal D}[\sigma] W_\sigma \Phi_\sigma\langle O\rangle_\sigma\over
\int {\cal D}[\sigma]W_\sigma \Phi_\sigma}\;,
\end{eqnarray}
where $\Tr_A$ denotes a trace at a fixed number of particles $A$.
$W_\sigma=G_\sigma|\Tr_A U_\sigma|$ is a positive-definite weight function,
$\Phi_\sigma=\Tr_A U_\sigma/|\Tr_A U_\sigma|$ is the Monte carlo ``sign'' and
 $\langle O\rangle_\sigma=\Tr_A [
OU_\sigma]/\Tr_A U_\sigma$. Both $\Tr_A U_\sigma$ and $\langle O\rangle_\sigma$ can be evaluated using matrix algebra in the single-particle space. For example, the
grand canonical trace of the one-body propagator $U_\sigma$ is
given by
\be\label{Tr-U-sigma}
\Tr\, U_\sigma=\det \left(1+{\bf U}_\sigma \right) \;,
\ee
where ${\bf U}_\sigma$ is the $N_s\times N_s$ matrix
representing $U_\sigma$ in the single-particle space containing $N_s$ single-particle orbitals.

The canonical trace can be evaluated using particle-number projection with  $\phi_m=2\pi m/N_s$ as quadrature points. For example, the canonical trace of $U_\sigma$ is given by
\be
\Tr_A U_\sigma= {1\over
N_s}\sum_{m=1}^{N_s} e^{-i\phi_m A}
\det\left(1+e^{i\phi_m}{\bf U}_\sigma\right)\;.
\ee

The multi-dimensional integral over the auxiliary fields in (\ref{eq:HS}) is evaluated by Monte Carlo methods. The auxiliary fields are sampled according to the distribution $W_\sigma$. For samples $\{\sigma_i\}$ the expectation value in (\ref{eq:obs}) is estimated from
\begin{eqnarray}
\label{eq:obsmc}\langle O\rangle \approx {\sum_i
\Phi_{\sigma_i}\langle O\rangle_{\sigma_i} \over\sum_i
\Phi_{\sigma_i}} \;.
\end{eqnarray}

Such auxiliary field  Monte Carlo (AFMC) methods have been developed in the framework of the interacting nuclear shell model~\cite{smmc}. We have recently extended AFMC to nanoparticles~\cite{afmc-nano}.

For a general interaction, the sign $\Phi_\sigma$ can fluctuate from sample to sample. When the statistical error of the sign is larger than its average value, the method breaks down. This is known as the Monte carlo sign problem. In the nuclear case, the dominating collective components of the interaction have a good sign. Interactions with small bad-sign components can be treated by the method of Ref.~\cite{alhassid94}.

\section{NUCLEI}\label{nuclei}

We have used AFMC to calculate statistical properties of nuclei in the iron region within the complete $fpg_{9/2}$ shell~\cite{smmc}. The single-particle Hamiltonian corresponds to a Woods-Saxon potential $V$ plus a spin-orbit interaction~\cite{BM69}. The two-body interaction~\cite{ABDK96,NA97} includes a monopole pairing interaction whose coupling strength is determined from the experimental odd-even mass differences. Also included are  multipole-multipole interactions that are obtained by expanding the surface-peaked interaction $v({\bf r}, {\bf r}^\prime)
 = -\chi (dV/dr)(dV/dr^\prime)\delta(\hat{\bf r} - \hat{\bf r}^\prime)$ into quadrupole, octupole and hexadecupole components. The coupling constant $\chi$  is determined self-consistently~\cite{ABDK96} $\chi^{-1} = \int_0^\infty dr \; r^2  \left(dV/ dr \right)  \left(d\rho/ dr
\right)$ ($\rho$ is the nuclear density). The quadrupole, octupole and hexadecupole interactions are then renormalized by  $2,1.5$ and $1$, respectively.   All interaction components are attractive and lead to a good-sign Hamiltonian.

In the following subsections we discuss signatures of pairing correlations in the heat capacity and moment of inertia of nuclei.

\subsection{Heat capacity}

The heat capacity $C=dE/dT$ is calculated by a numerical derivative of the thermal energy. We used the method of Ref.~\cite{la01}, in which the thermal energy at $\beta\pm\delta\beta$ are calculated with the same Monte Carlo walk and correlated errors are taken into account to estimate the statistical error of the numerical derivative.

\begin{figure}[h!]
\centering
\includegraphics[height=.37\textheight]{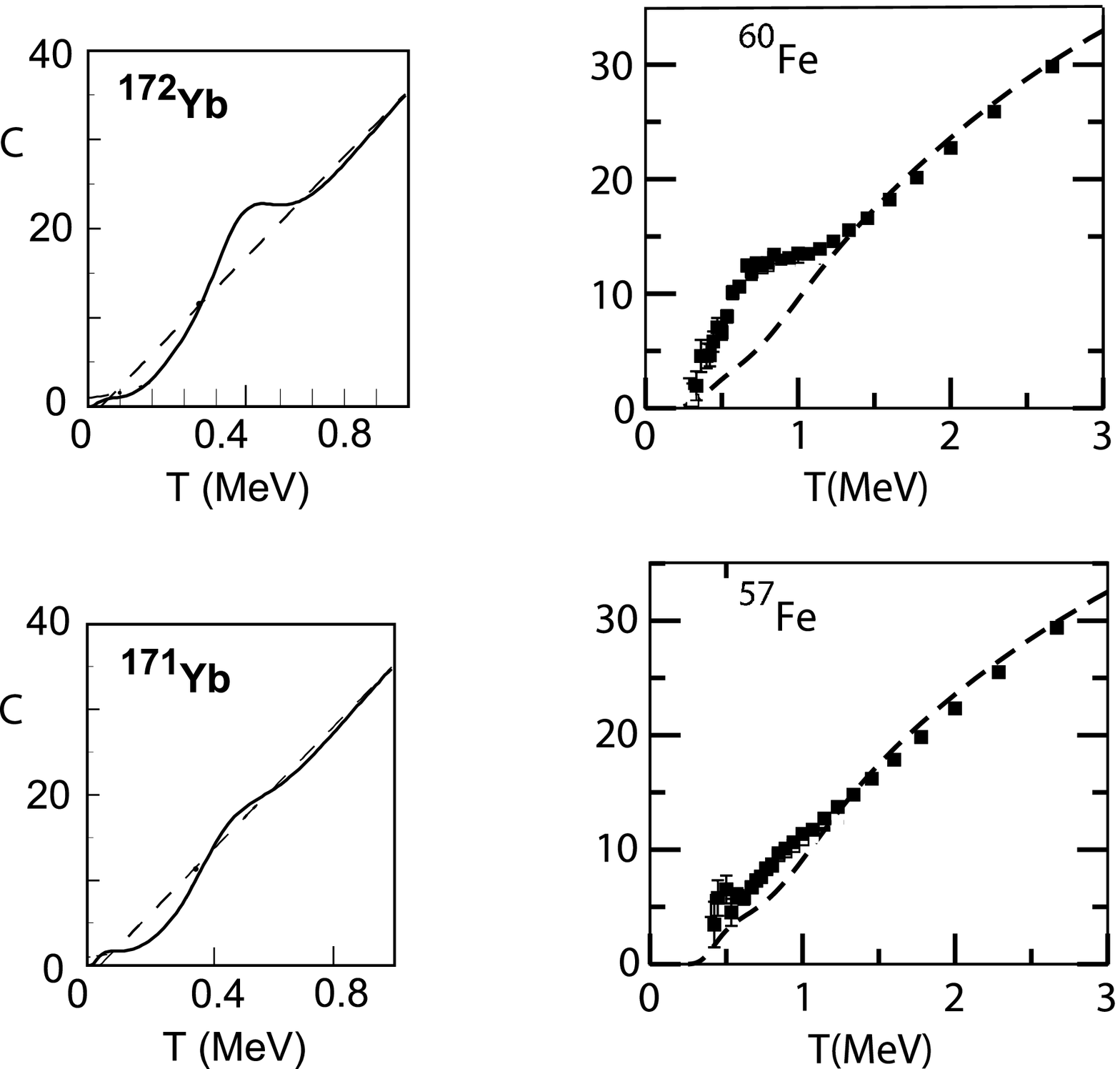}
\caption{Right: Heat capacities of $^{60}$Fe (top) and $^{57}$Fe (bottom) versus temperature $T$~\cite{alhassid03}. The solid squares describe the extended heat capacities while the dashed lines describe the heat capacities in the independent-particle model. Left: measured heat capacities in $^{172}$Yb  and $^{171}$Yb (solid lines) are compared with the heat capacities in the Fermi gas model (dashed lines)~\cite{schiller01}.}
\label{hc-nuclei}
\end{figure}

The Monte Carlo approach in a single major shell is valid for temperatures up to $T \sim 1.5-2$ MeV. At higher temperatures the heat capacity saturates because of truncation effects and it is  necessary to include higher shells. Monte carlo calculations in larger spaces are possible but may be time consuming.  Instead, we have combined the fully correlated calculations in the  truncated space with independent-particle model calculations in the full space (including all bound states and the continuum)~\cite{alhassid03}.

The right panels of Fig.~\ref{hc-nuclei} show the heat capacity versus temperature in the even-even nucleus $^{60}$Fe and even-odd nucleus  $^{57}$Fe. The solid squares describe the extended heat capacity, calculated by the method discussed above. These heat capacities have the correct approximate linear behavior at higher temperatures. The dashed lines in Fig.~\ref{hc-nuclei} are the results of the independent-particle model. In the even-even nucleus we observe an enhancement in the heat capacity (an $S$ shape curve), while in the even-odd nucleus the heat capacity remains close to the heat capacity of the independent-particle model. Similar even-odd effects were measured in the heat capacity of rare-earth nuclei~\cite{schiller01} as is demonstrated in the left panels of Fig.~\ref{hc-nuclei}.

\subsection{Moment of inertia}

The moment of inertia $I$ at finite temperature determines the spin distribution of energy levels in the framework of the spin cut-off model. It describes the response of the nucleus to rotations. For a rotationally-invariant Hamiltonian, $I$ is given by $I=\beta \langle \hat J_z^2\rangle$, where $\hat J_z$ is the $z$-projection of total angular momentum.

\begin{figure}[bth]
\centering
\includegraphics[height=.4\textheight]{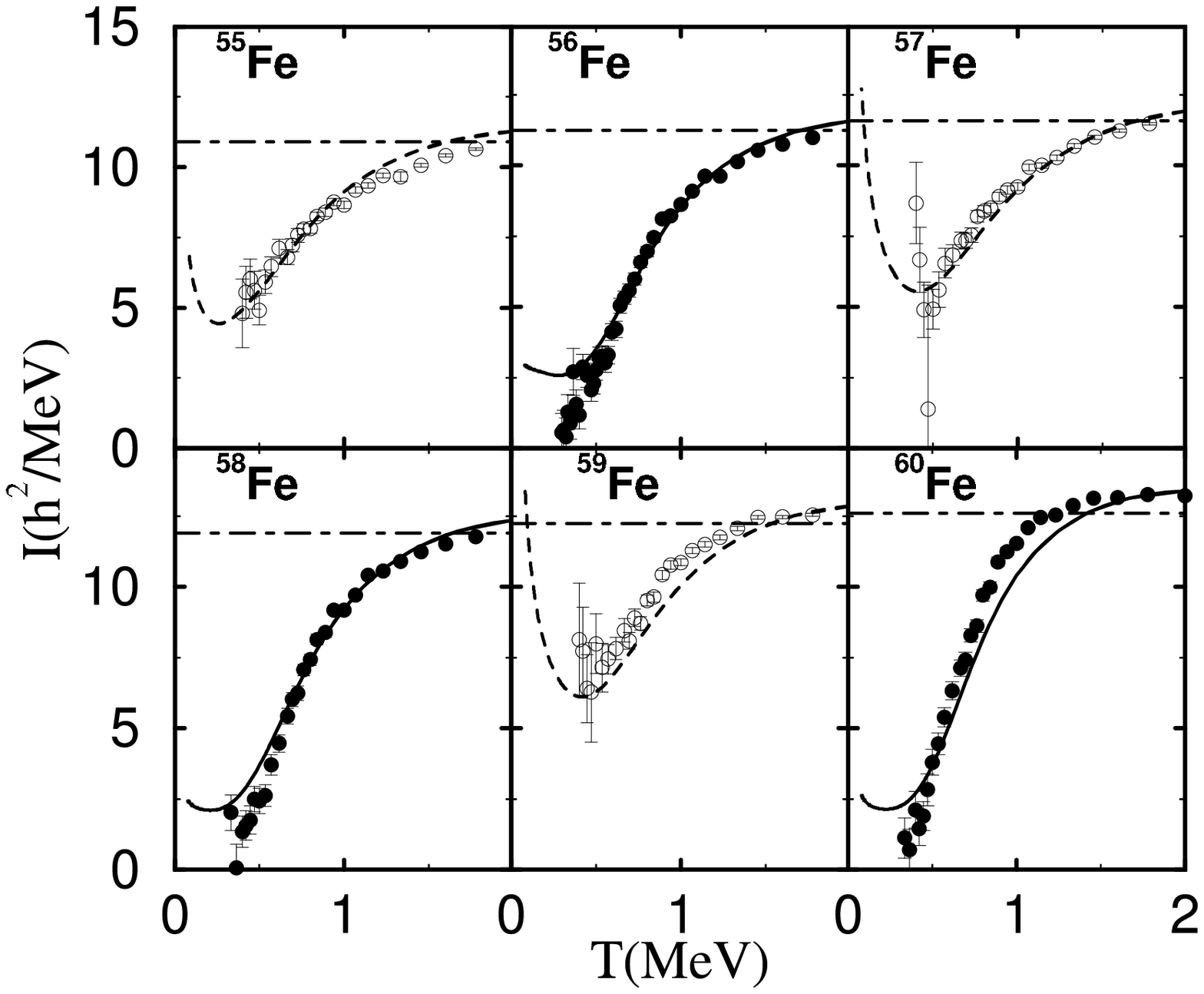}
\caption{
The moment of inertia versus $T$ for a family of iron
isotopes. The circles are the SMMC results, while the lines are the results of a simple model (see text). The solid lines are the
even number-parity projection while the dashed lines describe the odd
number-parity projection (for neutrons). The dotted-dashed lines are the rigid-body
moment of inertia.
The deformation parameter is $\beta_2=0.14$ and the pairing strengths are $g_p=0.42$ and $g_n=0.36$ for protons and neutrons, respectively. From Ref.~\cite{alhassid05}.
}
\label{inertia-T}
\end{figure}

Figure~\ref{inertia-T} shows the SMMC moment of inertia (symbols) as a function of $\beta$ for a series of iron isotopes. In general, we observe a suppression of the moment of inertia at low temperatures because of pairing correlations. In the even-odd isotopes, this suppression is weaker and we even observe an enhancement in the limit $T\to 0$.

This behavior of the moment of inertia can be explained in a simple model of a deformed Woods-Saxon potential plus a monopole pairing interaction~\cite{alhassid05}.  The single-particle states are organized in time-reversed pairs $k$ and $\bar k$ (corresponding to angular momentum projection $m$ and $-m$ along the symmetry axis). The simple Hamiltonian is then given by
\be\label{Hamiltonian}
H_{\rm def} = \sum_{k>0} \epsilon_k(a^\dagger_k a_k + a^\dagger_{\bar k} a_{\bar k}) - g P^\dagger P \;,
\ee
where
$P^\dagger=\sum_{k>0} a^\dagger_k a^\dagger_{\bar k}$ is the pair creation operator and $g$ is the pairing strength.

 The major odd-even effects are captured by a number-parity projection
\be
P_\eta=\frac{1}{2}(1+\eta e^{i \pi \hat A})\;,
\ee
where $\hat A$ is the particle-number operator. $P_\eta$ projects on even (odd) number of particles for $\eta=1$ ($\eta=-1$).

For a deformed nucleus, rotational symmetry is broken and the moment of inertia is a tensor $I_{ij}$. The number-parity projected intrinsic moment of inertia is given by
\begin{equation}\label{NP-inertia-xi}
{\cal I}_{ij}^\eta={\int_0^\beta d\tau\;\langle J_i(\tau)J_j(0)\rangle
+ \eta \prod_{k>0}\tanh^2 {\beta E_k \over 2}
\int_0^\beta d\tau\;\langle J_i(\tau)J_j(0)\rangle_\pi
\over 1+\eta \prod_{k>0} \tanh^2 {\beta E_k \over 2}} \;,
\end{equation}
where $J_i(\tau)$ describes the $i$-th component of the angular momentum in the intrinsic frame at imaginary time $\tau$.  The expression for
$\int_0^\beta d\tau\,\langle J_i(\tau)J_j(0)\rangle_\pi$ is obtained
from the expression
for $\int_0^\beta d\tau\,\langle J_i(\tau)J_j(0)\rangle$ (which depends on the quasi-particle occupations $f_k={1\over 1+e^{\beta E_k}}$) by the substitution
\be
f_k \rightarrow \tilde{f_k}={1\over 1-e^{\beta E_k}}\;.
\ee

Fluctuations in the pairing order parameter (i.e., the gap) are included in the static path approximation (SPA).  Finally, rotational symmetry is restored by integrating over all orientations of the intrinsic frame.  The results of the simple model are shown by the lines in Fig.~\ref{inertia-T}.

\section{NANOPARTICLES}\label{nanoparticles}

The spectra of ultra-small metallic grains (nanoparticles) of size $\sim 2-10$ nm were determined in Refs.~\cite{nano-experiments} by connecting them to leads and measuring their non-linear conductance. The number of electrons on the grain is controlled by varying a gate voltage.

The Hamiltonian of the metallic grain is given by
\begin{equation}\label{nanograin}
H_{\rm nanoparticle} = \sum_{\lambda \sigma} \left(\epsilon_\lambda - e\alpha V_g +
{1 \over 2} g_B \mu_B B \sigma \right) a^\dagger_{\lambda \sigma} a_{\lambda \sigma} + {e^2 \hat N^2 \over 2 C} - g \sum_{
\lambda,\mu} a_{\lambda +}^\dagger
a_{\lambda-}^\dagger a_{\mu-}a_{\mu+}\;.
\end{equation}
The first term on the r.h.s.~is a one-body Hamiltonian describing electrons in spin-degenerate orbital levels $\lambda$ ($\sigma=\pm$ corresponds to spin up/down electrons). Also included is a Zeeman term describing the coupling to an external magnetic field $B$ ($\mu_B$ is the Bohr magneton and $g_B$ is the $g$-factor).  $e^2 \hat N^2 / 2 C$ is a charging energy term that describes the Coulomb energy of the grain with capacitance $C$. The last term on the r.h.s.~is a reduced BCS pairing interaction describing the scattering spin up/down electron pairs between different orbital levels.

The Hamiltonian (\ref{nanograin}) can be solved in the BCS approximation. Below a certain critical temperature $T_c$, a superconducting solution exists. BCS theory is a mean-field approximation and is valid in the limit when the pairing gap $\Delta$ is large compared with the single-particle mean-level spacing $\delta$, .i.e., $\Delta/\delta \gg 1$. This holds for the larger grains and a pairing gap was observed in the excitation spectrum of a grain with an even number of electrons~\cite{nano-experiments}.

However, for the smaller grains the ratio $\Delta/\delta$ is comparable or smaller than $1$. This is the crossover to the fluctuation-dominated regime where BCS theory breaks down. Static fluctuations in the gap order parameter can be taken onto account in the SPA and small amplitude quantal fluctuations are included in the SPA plus RPA approach~\cite{lauritzen93}. However, at lower temperatures it is necessary to account for additional quantal fluctuations. For small number of single-particle levels and electrons, diagonalization methods were used to find the eigenvalues~\cite{mastellone98}. For the pairing Hamiltonian, it is possible to reduce the eigenvalue problem to a set of non-linear equations~\cite{richardson}, and this method was used in Ref.~\cite{lorenzo00}. However, at higher temperatures the number of relevant many-body levels increases rapidly and the method becomes impractical.

We have recently extended the AFMC methods to nanoparticles~\cite{afmc-nano}. The attractive pairing interaction is a good-sign interaction in a density decomposition and thus accurate Monte Carlo calculations are possible. The AFMC method can be used for a large number of single-particle levels and are valid at both low and high temperatures. A different Monte Carlo method that is suitable for pairing-type interactions was also applied recently to nanoparticles~\cite{vanhoucke06}.

\subsection{AFMC for nanoparticles}

In the following we consider a grain with a fixed number of electrons in the absence of magnetic field. The charging energy becomes a constant and can be ignored. The Hamiltonian (\ref{nanograin}) can be written in a density decomposition
\begin{eqnarray}\label{density-H}
H=\sum_\lambda\epsilon_\lambda \hat n_\lambda-{g\over 4}\sum_{\lambda\mu}
[(\rho_{\lambda\mu}+\bar{\rho}_{\lambda\mu})^2\ -(\rho_{\lambda\mu}-\bar{\rho}_{\lambda\mu})^2] \;,
\end{eqnarray}
where $\hat n_{\lambda}=a_{\lambda +}^\dagger a_{\lambda +}+a_{\lambda -}^\dagger a_{\lambda -}$ is the number operator of level $\lambda$, $\rho_{\lambda\mu}=a_{\lambda +}^\dagger a_{\mu +}$ are spin-up density operators and $\bar{\rho}_{\lambda\mu}=a_{\lambda -}^\dagger a_{\mu -}$  is the time-reversed operator of $\rho_{\lambda\mu}$. The interaction in (\ref{density-H}) has a quadratic form and is thus suitable for an HS decomposition (\ref{eq:HS}).

In a metallic grain, the band width is determined by the Debye frequency $\omega_D$. In an $Al$ grain, the measured Debye frequency is $\omega_D\approx 34$ meV. For the gap, we use the experimental value of thin films $\Delta\approx 0.38$ meV. The coupling constant is determined from $\Delta \approx 2\omega_D e^{-\delta/g}$ to be $g/\delta\approx 0.193$. In a grain with $\Delta/\delta=1$, the number of single-particle levels is then $\approx 180$.

In practical AFMC calculations, we truncate the model space to a smaller band width. This can be done by renormalizing the coupling strength $g$. A simple way of determining the new coupling constant in the truncated space is by keeping the BCS gap (for a discrete single-particle spectrum) fixed. For a picketfence (i.e., equally spaced) spectrum, the renormalized constant $g_r$ in a truncated space with $2N_r+1$ single-particle levels and half filling is given approximately by
\be
{g_r \over \delta} = {1 \over {\rm arcsinh}\left({N_r+1/2 \over \Delta/\delta}\right)} \;.
\ee

The choice of the truncated model space depends on the temperature range of interest and higher temperatures require larger model spaces. A model space of $N_r=25$ is good for temperatures up to $T \sim 3 \delta$. The AFMC calculations scales as $\sim N_r^4$ and at lower temperatures we can increase the efficiency of our calculations by truncating the model space further. We used $N_r=15$ for $T\leq 1.5\,\delta$ and $N_r =10$ for $T\leq 0.8\,\delta$.

\subsection{Signatures of pairing correlations}

We used AFMC to calculate  the heat capacity and spin susceptibility as a function of temperature for both even and odd number of electrons. Since the imaginary time is discretized, it is necessary to correct for systematic errors in the size of the time slice $\Delta \beta$. We  calculated the quantities of interest for two time slices $\Delta\beta=1/32$ and $\Delta\beta=1/64$, and then extrapolated the results to $\Delta\beta=0$.

\begin{figure}[bth]
\includegraphics[height=.26\textheight]{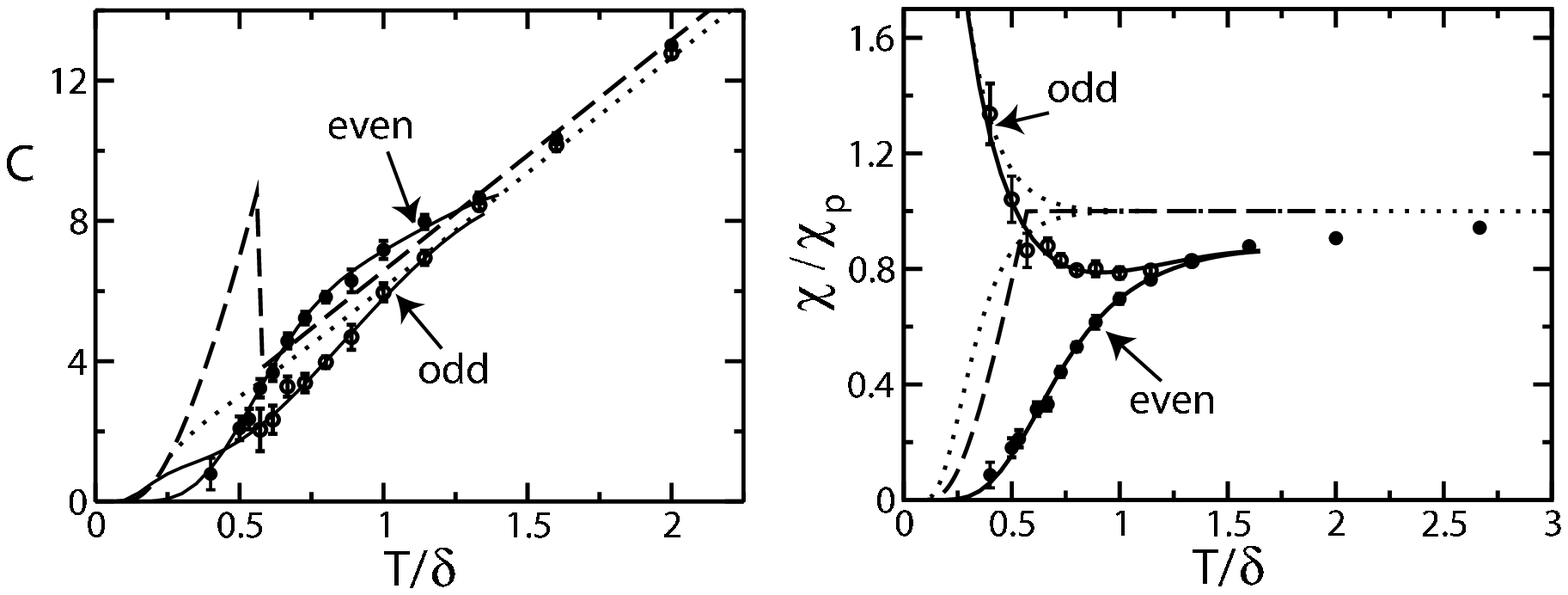}
\caption{\label{hc-chi-nano}
Left: the heat capacity of an ultra-small metallic grain with $\Delta/\delta=1$ versus temperature $T$. Solid (open) circles are the AFMC results for a grain with an even (odd) number of electrons. The results are compared with the BCS approximation (dashed line) and the canonical independent-particle model (dotted). Right: the spin susceptibility $\chi$ (in units of $\chi_p$) versus $T$ for a grain with $\Delta/\delta=1$. Symbols and lines are as in the left panel. After Ref.~\cite{afmc-nano}.
}
\label{hc-chi-nano}
\end{figure}

The heat capacity is calculated using the same method as in the nuclear case~\cite{la01}.  The left panel of Fig.~\ref{hc-chi-nano} shows the heat capacity versus $T$ for a grain with $\Delta/\delta=1$. The AFMC calculations for even (odd) number of electrons are shown by solid (open) circles.  The dashed line is the BCS result, for which the heat capacity displays a discontinuity at the critical temperature. In the AFMC calculations, the heat capacity is a smooth function of temperature, but for an even number of electrons we observe an enhancement of the heat capacity similar to the $S$-shape heat capacity observed in even-even nuclei.  At higher temperatures, the heat capacity approaches the Fermi gas result $C= 2 \pi^2 T/3\delta$ valid for $T \gg \delta$ (but still much below the Fermi energy).

The spin susceptibility $\chi=d M/dB |_{B=0}$  describes the magnetic response of the grain (i.e., magnetization $M$) to a weak magnetic field $B$. It is given by
\be
\chi= g_B^2 \mu_B^2 \beta \left( \langle \hat S_z^2\rangle - \langle \hat S_z\rangle^2 \right)\;,
\ee
where $g_B$ is the $g$-factor of electrons in the grain, $\mu_b$ is the Bohr magneton and $\hat S_z$ is the $z$ component of the total spin of the grain. The right panel of Fig.~\ref{hc-chi-nano} shows the AFMC spin susceptibility versus temperature for both even (solid circles) and odd (open circles) number of electrons. The dashed line is the BCS result and the dotted lines are the canonical Fermi gas results. In comparison with the canonical Fermi gas results, we observe that pairing correlations suppress the spin susceptibility and even-odd effects persist to higher temperatures. The susceptibility for an odd number of particles diverges in the limit $T \to 0$ already in the Fermi gas picture (i.e., without a  pairing interaction).  This is a result of the finite spin of the odd particle occupying the orbital at the Fermi energy. However, in the presence of pairing correlations, the spin susceptibility for an odd number of electrons initially decreased with decreasing temperatures, displaying a minimum at a finite temperature before diverging at $T \to 0$. For large temperatures the spin susceptibility approaches the value $\chi_p= g_B ^2\mu_B^2/2\delta$, obtained for a Fermi gas at $T\gg \delta$.

The above results for the heat capacity and spin susceptibility of nanoparticles demonstrate that pairing correlations in the fluctuation-dominated regime manifest through effects that depend on the particle-number parity~\cite{vondelft01}.

\section{CONCLUSION}

We discussed pairing correlation in nuclei and nanoparticles in the crossover between the bulk BCS limit and the fluctuation-dominated regime.  Auxiliary-field Monte Carlo methods are used to include fluctuations beyond the BCS mean-field theory. The heat capacity and moment of inertia of nuclei exhibit pairing correlation that depend on the particle-number parity of protons and neutrons. Similar even-odd effects (in the number of electrons) are observed in the heat capacity and spin susceptibility of nanoparticles.

\section*{Acknowledgments}
I  would like to thank  G.F. Bertsch, L. Fang, S. Liu and S. Schmidt for their collaboration on the work presented above.


\begin{thebibliography}{9}

\bibitem{BCS} J. Bardeen, L.N. Cooper and J.R. Schrieffer,
Phys. Rev. {\bf 108} (1957) 1175.

\bibitem{vondelft01}
J.von Delft and D.C. Ralph, Phys. Rep. {\bf 345} (2001) 661.

\bibitem{lang93}
G.H. Lang, C.W. Johnson, S.E. Koonin, and W.E. Ormand,
Phys. Rev. C {\bf 48} (1993) 1518.

\bibitem{alhassid94} Y. Alhassid, D.J. Dean,
S.E. Koonin, G.H. Lang, and W.E. Ormand, Phys. Rev.
Lett. {\bf 72} (1994) 613.

\bibitem{afmc-nano}
Y. Alhassid, L. Fang and S. Schmidt, to be published.

\bibitem{la01} S. Liu and Y. Alhassid, Phys, Rev. Lett. {\bf 87}  (2001) 022501.

\bibitem{alhassid03} Y. Alhassid, G.F. Bertsch and L. Fang, Phys. Rev. C {\bf 68} (2003) 044322.

\bibitem{alhassid05} Y. Alhassid, G.F. Bertsch, L. Fang and S. Liu, Phys. Rev. C {\bf 72} (2005) 064326.

\bibitem{HS}
J. Hubbard, Phys. Rev.Lett. {\bf 3} (1959) 77; R.L.
Stratonovich, Dokl. Akad. Nauk. S.S.S.R. {\bf 115} (1957) 1097.

\bibitem{smmc} For a recent review, see  Y. Alhassid, Int. J. Mod. Phys.
B {\bf 15} (2001) 1447.

\bibitem{BM69} A. Bohr and B. R. Mottelson, {\em Nuclear Structure}
vol. 1, Benjamin, New York, 1969.

\bibitem{ABDK96} Y. Alhassid, G.F. Bertsch, D.J. Dean and S.E. Koonin,
Phys. Rev. Lett.  {\bf 77} (1996) 1444.

\bibitem{NA97} H. Nakada and Y. Alhassid,
Phys. Rev. Lett. {\bf 79} (1997) 2939.

\bibitem{schiller01}  A. Schiller, A. Bjerve, M. Guttormsen, M. Hjorth-Jensen, F. Ingebretsen,  E. Melby, S. Messelt, J. Rekstad, S. Siem, and S.W. Odegard, Phys. Rev. C {\bf 63}  (2001) 021306.

\bibitem{nano-experiments}
D.C. Ralph, C.T. Black, and M. Tinkham, Phys. Rev. Lett
{\bf 74} (1995) 3241; {\em ibid.} {\bf 78} (1997) 4087.

\bibitem{lauritzen93}
B. Lauritzen, M. Anselmino, P.F. Bortignon and R.A. Broglia, Ann. Phys. {\bf 223} (1993) 216.

\bibitem{mastellone98}
A. Mastellone, G. Falci and R. Fazio, Phys. Rev. Lett. {\bf 80} (1998) 4542.

\bibitem{richardson}
R.W. Richardson, Phys. Rev. Lett. {\bf 3} (1963) 277; Phys. Rev. {\bf 159} (1967) 792.

\bibitem{lorenzo00}
A. Di Lorenzo, R. Fazio, F.W.J. Hekking, G. Falci, A.
Mastellone, and G. Giaquinta, Phys.\ Rev.\ Lett {\bf 84} (2000) 550.

\bibitem{vanhoucke06}
K. van Houcke, S.M. A. Rombouts, L. Pollet, arXiv:
cond-mat/0603541.

\end{thebibliography}
\end{document}